\documentclass[12pt]{iopart}
\usepackage{ae}
\usepackage{epsf,epsfig}
\usepackage{graphicx}
\usepackage{amssymb,dcolumn,bm}

\begin{document}

\title
{Asymmetric exclusion processes with shuffled dynamics}
\author{Marko W\"olki\dag\footnote[3]{woelki@traf1.uni-duisburg.de}, 
Andreas Schadschneider\ddag\footnote[4]{as@thp.uni-koeln.de}, 
and Michael 
Schreckenberg\dag\footnote[5]{schreckenberg@uni-duisburg.de}
} 

\address{\dag Theoretische Physik, Fachbereich Physik, Universit\"at 
Duisburg-Essen, Lotharstr. 1, D-47057 Duisburg, Germany} 
\address{\ddag Institut f\"ur Theoretische Physik, Universit\"at zu
K\"oln, Z\"ulpicher Str. 77, D-50937 K\"oln, Germany } 

\date{\today}

\begin{abstract}
  The asymmetric simple exclusion process (ASEP) with periodic
  boundary conditions is investigated for shuffled dynamics. In this
  type of update, in each discrete timestep the particles are updated
  in a random sequence. Such an update is important for several
  applications, e.g.\ for certain models of pedestrian flow in two
  dimensions.  For the ASEP with shuffled dynamics and a related
  truncated process exact results are obtained for deterministic
  motion ($p=1$). Since the shuffled dynamics is intrinsically
  stochastic, also this case is nontrivial. For the case of stochastic
  motion ($0<p<1$) it is shown that, in contrast to all other updates
  studied previously, the ASEP with shuffled update does not have a
  product measure steady state. Approximative formulas for the steady
  state distribution and fundamental diagram are derived that are in
  very good agreement with simulation data.
\end{abstract}

\maketitle






\section{Introduction}

The asymmetric simple exclusion process (ASEP) is one of the most
studied models far from equilibrium \cite{derrida,schuetz,cssrev}. It
has been used to describe various problems in many fields of research,
such as biopolymerization and traffic flow. A lot of analytical
results exist, both for open and periodic boundary conditions. The
model describes a particle system on a chain with hard core exclusion.
Particles are allowed to hop one site to their right, supposed that it
is empty. The usual dynamics is in continuous time: the
random-sequential update \cite{derrida}. Moreover, discrete-time
update schemes have been studied: backward- and forward-ordered
sequential updates, site-oriented \cite{rajewskiseq} as well as
particle-oriented \cite{evansseq}, sublattice-parallel updates
\cite{honecker,hinrichsen} and fully-parallel dynamics
\cite{ito,speer,nienhuis} (for an overview, see \cite{rajewski}). We
analyze another update scheme, that has originally been introduced in
a two-dimensional cellular automaton describing pedestrian dynamics
\cite{kluepfel,kluepfeldiss}.

A typical situation encountered in pedestrian dynamics is the motion
along a corridor. This is modeled as a strip of finite width $W$ in
$y$-direction of a cartesian coordinate plane and length $L$ in
$x$-direction divided into square cells. Each cell can be in one of
two possible states, i.e. either occupied by one of the $N$
pedestrians, or empty (hard core exclusion rule). Assuming periodic
boundary conditions in $x$-direction and impenetrable walls in
$y$-direction implies that the density of pedestrians $\rho = N/(W
\cdot L)$ is constant. A pedestrian can move to neighbor cells with
different transition probabilities, depending on the direction and the
particular neighborhood. The preferred direction is along the
$x$-axis. A parallel update, in which all particles are updated
simultaneously, would lead to conflicts, in which more than one person
tries to access the same cell \cite{friction}. Hence, another way of
updating was chosen, the so-called shuffled dynamics: At each
(discrete) timestep, the order in which particles are allowed to move
is determined by a random permutation. Since the theoretical
implications of this update procedure have not been considered before,
we here investigate the one-dimensional limit first. In the language
of pedestrian dynamics this corresponds to a very narrow corridor with
$W=1$ such that `side-by-side' motion or overtaking is not
possible. The model then becomes equivalent to the ASEP with shuffled
update.

The difference between this update scheme and the random-sequential
update mentioned above is analogous to an urn problem with and without
replacement: If one imagines an urn with $N$ balls, numbered $1, 2,
\dots, N$, the random-sequential update can be realized by choosing a
ball at random, updating the particle with the ball-number and
replacing the ball into the urn afterwards. In the shuffled update,
one chooses a ball and updates the corresponding particle without
replacing the ball. Then, one chooses the next ball, and so on, until
the last particle is updated. Then the urn is refilled and the
procedure is repeated in the next timestep.

After giving a precise definition of the model, approximative formulas
for the steady state distribution and fundamental diagram are
derived. By mapping onto generalized zero-range processes it is shown
that the ASEP with shuffled update does not have a product measure
steady state for general parameter choice. Some applications and
generalizations are given afterwards and concluding the results are
discussed and some interesting directions for further research are
given.


\section{ASEP with shuffled dynamics}

In the following we give a more formal definition of the ASEP with
shuffled dynamics. Consider a one-dimensional lattice with $L$ sites
and periodic boundary conditions. Each site may either be occupied by
one of the $N$ particles, labelled $i={1, 2, ..., N}$, or it may be
empty. Therefore the particles are distinguishable. In each discrete
timestep a random permutation $\pi(1, ..., N)$ of the particle labels
equals the update sequence. If the right neighboring cell is empty,
the relevant particle moves one site to the right with probability
$p$, if it is occupied, the particle stays in its cell.

\begin{figure}[!hbt]
\begin{center}
\epsfig{figure=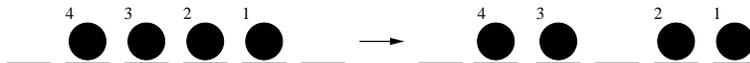,width=10.0cm,clip=}
\end{center}
\caption{\label{shuffbild}Shuffled update of a cluster consisting of
four particles numbered from right to left. The drawn sequence
is $3,4,1,2$ and $p=1$.}
\end{figure}

Figure \ref{shuffbild} shows a part of a large system consisting of
six cells and four particles (numbered $1, 2, 3, 4$ without loss of
generality), at time $t$ (left) and $t+1$ (right). The drawn update
sequence is $\dots, 3, \dots, 4, \dots, 1, \dots, 2, \dots$, where the
ellipsis indicate that other particle numbers belonging to different
clusters (units of neighboring particles) can be chosen between (for
the cluster depicted, only the relative positions of the numbers of its
particles in the sequence are of interest). Particle 3, chosen first,
can not move since the cell in front is occupied by 2, and similar for
particle 4. Considering the case $p=1$, particle 1 then moves
deterministically to the right. Then 2 also moves, because it was
drawn after 1. Although particle 4 is drawn after 3, it can not move,
since both were drawn before 1 and 2.

As pointed out in the Introduction, the shuffled update is different
from the random-sequential dynamics which is generically used for the
ASEP. Whereas the latter describes stochastic processes in continuous
time, the shuffled dynamics combines elements of discrete updates and
dynamics in continuous time. It is discrete in the sense that there is
a well-defined timestep during which each particle is updated exactly
once.  On the other hand, the order of updating the particles is not
fixed. E.g.\ it may happen that a specific particle is updated last
during a timestep and first during the next one! This can actually be
a problem for applications because it is difficult to identify the
timestep as a kind of reaction time (as it is natural for the case of
parallel dynamics \cite{ito}).

Despite this important difference, the two updates share certain
similarities. In contrast to the ordered updates with fixed order,
they do not have a deterministic limit, even for hopping probabilities
$p=1$. However, in the random-sequential case the dynamics depends on
$p$ only in a trivial way, since by rescaling time always $p=1$ can be
chosen. This is not possible in the shuffled case. We therefore expect
a non-trivial $p$-dependence of the results, as in the other
discrete-time updates.


\section{Steady State Distribution}
\label{COMFsection}

Using a particle-oriented representation \cite{evansseq,ito}, the
state of a system with $N$ particles at time $t$ can uniquely be
specified by $|n_{1}(t)^{(\pi_{1}(t))},$
$\dots,n_{N}(t)^{(\pi_{N}(t))}\rangle$.  Here $n_{j}$ represents the
state of the $j$-th particle (i.e. the number of empty sites in front)
and $\pi_{j}$ its update number (the position of the $j$-th particle
in the ordered update sequence). At each timestep, the permutation
operator $\mathcal{P}(t)$ generates the new random sequence as
follows: $\mathcal{P}(t)|\dots, n_{i}(t)^{(\pi_{i}(t))},\dots,\rangle
= |\dots,n_{i}(t)^{(\pi_{i}(t+1))},\dots\rangle$, i.e. without
changing the variables $n_{i}(t)$. The complete update of the system
$t \rightarrow t+1$ can be described by the action of the
transfer-matrix $h^{(1)}h^{(2)}\dots h^{(N)}$, being an ordered
product of the local operators $h^{(j)}$ acting on the state of the
particles $\sigma_{j}-1$ and $\sigma_{j}$: $h^{(j)} =
h^{(j)}_{\sigma_{j}-1, \sigma_{j}} = {\bf 1} - p
(a_{\sigma_{j}}a_{\sigma_{j}}^{\dagger}
-a_{\sigma_{j}-1}^{\dagger}a_{\sigma_{j}})$, where $\sigma_{j}$ is the
particle with update number $j$. \footnote{For a detailed explanation
of the used formalism, see \cite{evansseq}.} The master equation reads
\begin{equation}
\label{masterP}
\fl \langle P | \mathcal{P}(t)\cdot \dots h^{(i)}\dots
|\dots,n_{i}(t)^{(\pi_{i}(t))},\dots \rangle = \langle
P|\dots,n_{i}(t+1)^{(\pi_{i}(t+1))},\dots\rangle,
\end{equation}
where $\langle P|\dots, n_{i}(t), \dots\rangle$ denotes the
probability of the configuration $|\dots, n_{i}(t), \dots\rangle$.  In
the steady state, equation (\ref{masterP}) simplifies, because all
time-dependencies vanish. Since the permutation operator generates
random sequences (each of them with probability $1/N!$) one can
calculate the probability with which a particle moves in a certain
configuration. The $n$-th particle of a cluster has moved at time
$t+1$, if and only if at least the first $n$ particles are chosen in
the order from the right to the left (with probability $1/n!$) and if
they all moved (with probability $p^{n}$). The probability that
exactly $l$ particles of a cluster of length $n$ (with $n>l$) move is
given as the probability for $l$ particles to move, as calculated
above, minus $p^{k+1}/(k+1)!$, the probability for the $(l+1)$-th
particle to move. We arrive at $u_{l}(n)$, the probability for $l$
particles leaving a cluster of length $n$ (where $0\leq l\leq n$):
\begin{equation}
\label{transitionshuffled}
u_{l}(n) = \frac{p^{l}}{l!} - \frac{p^{l+1}}{(l+1)!}\theta(n-l) 
\end{equation}
(and $0$ otherwise). The Heaviside step function ensures that in the
case $l=n$ the right term vanishes. Note that although the particles
in the ASEP are updated in $N$ steps, it can be considered as if
all clusters were updated in parallel.

 We now leave the operator notation and write the steady state
  probability simply as $P(n_{1}, n_{2}, \dots, n_{N})$. We assume
  that this probability can be written as a product of the
  probabilities $P_{n_{i}}$ to find particle $i$ with $n_{i}$ holes in
  front, i.e.
\begin{equation}
\label{fact}
P(n_{1}, n_{2}, \dots, n_{N}) = P_{n_{1}}P_{n_{2}}\dots P_{n_{N}}.
\end{equation}
This constitutes the so-called car-oriented mean-field (COMF) theory,
  successfully applied previously to traffic flow models
  \cite{schads,revisited}. Note that equation (\ref{fact}) implies
  that only correlations between neighboring particles are taken into
  account. In the thermodynamic limit, the master equation for the
  steady state reads:
\begin{eqnarray}
  \label{P00}
  P_{0} & = P_{0}-\left(p - g \right)(1-P_{0}) + p{\bar g} P_{1},\\
  \label{P01}
  P_{1} & = \left(p - g\right)(1-P_{0}) + \left(pg + {\bar
  p}{\bar g}\right) P_{1} + p{\bar g} P_{2},\\
  \label{P0n}
  P_{n} & = {\bar p}g P_{n-1} + \left(pg + {\bar p}{\bar g}\right) P_{n} +
  p{\bar g} P_{n+1}, & \forall n>1.
\end{eqnarray}
Here $P_{k}$ ($k=0, 1, \dots$) is the probability for an arbitrary
particle to have $k$ holes (a hole-cluster of length $k$) in front and
$g=1-{\bar g}$ is the probability that the particle in front moves.
As an example we explain the equation for $P_{1}$ in the following: A
gap of $1$ hole can arise from three different processes corresponding
to the three terms on the r.h.s. of (\ref{P01}): If the particle has
$2$ holes in front (probability $P_{2}$) the gap decreases by $1$ if
the particle itself moves (probability $p$), but the preceeding one
not (probability $\bar{g}$). The gap remains unchanged (probability
$P_{1}$) if either both particles move (probability $pg$) or if both
particles do not move (probability $\bar{p}\bar{g}$). If the particle
has its preceeding particle directly in front (probability $P_{0}$)
the situation is more sophisticated: The particle has one hole in
front afterwards if and only if the particle in front has moved, but
the particle itself not. This probability depends on the position of
the particle in the cluster to which itself and the preceeding
particle belong. The probability $P(k)$ for the particle to have
exactly $k-1$ particles directly in front is approximated by
\begin{equation}
\label{Pvonk}
P(k) = P_{0}^{k-1}(1-P_{0}).
\end{equation} 
This probability (\ref{Pvonk}) has to be multiplied with
$u_{k-1}(n)=\frac{p^{k-1}}{(k-1)!}-\frac{p^k}{k!}$, for $k \leq n$,
obtained from equation (\ref{transitionshuffled}). Finally we have to
sum over all possible $k$. This yields
$(1-P_{0})\sum_{k=2}^{\infty}\left(\frac{p^{k-1}}{(k-1)!}-\frac{p^{k}}{k!}
\right)P_{0}^{k-1}$ which can be rewritten as the first term on the
r.h.s. of (\ref{P01}) as one can check easily if one interpretes $g$
as the hopping probability for a particle, averaged over all possible
numbers of particles in front, i.e.
\begin{equation}
  \label{gshuffled}
    g = \sum\limits_{k=1}^{\infty}\frac{p^{k}}{k!}P(k) =
    \cases{ p,& for $P_{0} = 0$,\\ \frac{1-P_{0}}{P_{0}}
    \left(\exp(pP_{0})-1\right), & for $P_{0}>0$.}
\end{equation}
Using this, the system of equations (\ref{P00}) - (\ref{P0n}) can be
solved by generating functions \cite{schads,revisited}.  We obtain an
implicit expression for $P_{0}$ that has to be solved numerically in
general, since $g$ and $\bar{g}$ depend on $P_0$ via
(\ref{gshuffled}):
\begin{eqnarray}
  P_{0} &=& \frac{p(\rho-\bar{\rho}) -
  (p\rho-\bar{\rho})g}{p\rho\bar{g}},\label{P0allg}\\
 P_{n} &=&
  \frac{(p-g)(1-P_{0})}{\bar{p}g}\left(\frac{\bar{p}g}{p\bar{g}}\right)^n,
  \qquad  \forall n>0.
\label{Pnallg}
\end{eqnarray}
Again, the abbreviations $\bar{\rho} = 1- \rho$, etc.\ were used.
Since $g$ is the probability that an arbitrary particle moves, it
equals also the average velocity $\bar{v}$ of the particles. This is
related to the flow $J$ and yields the so-called fundamental diagram
\begin{equation}
  \label{gfunda}
  J(\rho) = \bar{v}\rho = g\rho.
\end{equation}

\begin{center}
\begin{figure}[!ht]
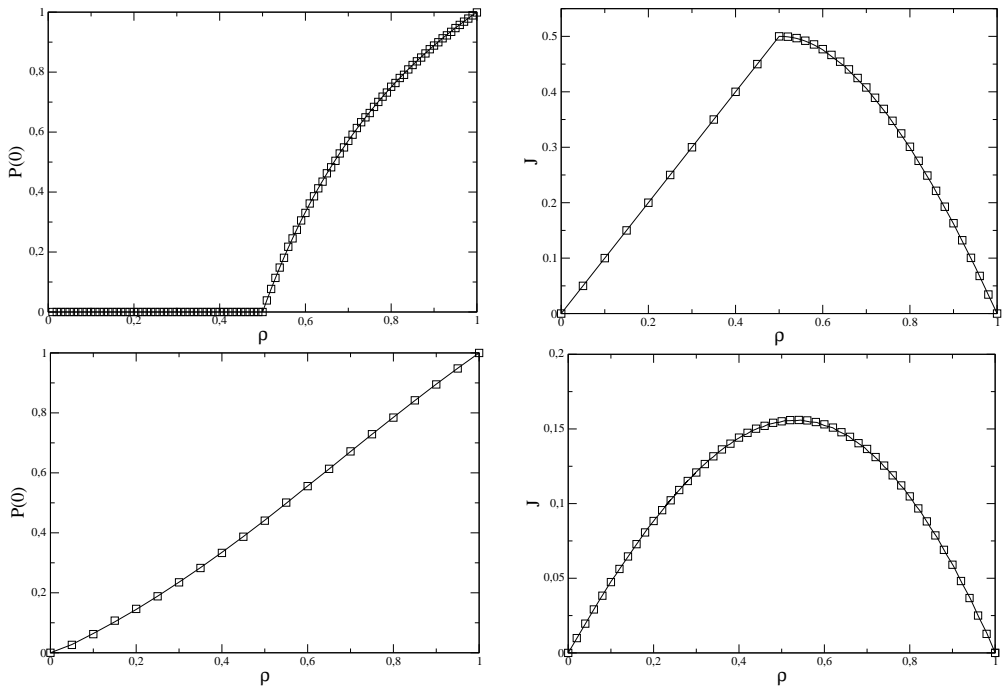

\centerline{\epsfig{figure=shuffp1P0.eps, height=4.5cm, clip=}
\quad\epsfig{figure=exp_hyper_det.eps, height=4.5cm, clip=}}
\centerline{\epsfig{figure=shuffp05P0.eps, height=4.5cm, clip=}
\quad\epsfig{figure=COMF_q_05.eps, height=4.5cm, clip=}}
  \caption{The probability $P_{0}(\rho, p)$ (left) and the fundamental
    diagram $J(\rho,p)$ (right) for $p=1$ (top) and $p=0.5$
    (bottom). Depicted are the results from COMF (lines)
    and from computer simulations (squares) for a system 
    consisting of $L=500$ cells. Note that $P_{0}(\rho,p=1)$ is exact.
    }
  \label{p1Bild}
\end{figure}
\end{center}

In the (partially deterministic) case $p=1$, $P_{0}$ becomes 
\begin{equation}
\label{P0}
P_{0} =\frac{2\rho-1}{\rho}\theta \left(\rho-\frac{1}{2}\right), 
\end{equation}
as one can check easily. Note that $p=1$ and $\rho \leq 1/2$ is the
only case in which $P_{0}$ can vanish. For $p=1$, the probability
$P_{0}$ is completely determined by the fact that the first particle
of a cluster moves deterministically and all the other particles have
smaller hopping probabilities due to the shuffling. For densities
$\rho \leq 1/2$ this implies that any state which consists only of
clusters of size 1, i.e.\ separated particles, is stationary. Hence
the probability to find a particle directly in front vanishes and we
have $P_{0}=0$ (Fig.~\ref{p1Bild}).  For densities $\rho > 1/2$
clusters are formed that are separated by exactly one hole, i.e.\
there are only isolated empty cells in the steady state. It is easy to
verify that no pairing of holes can happen, since from the point of
view of the holes they jump at least one site backwards and at most to
the end of the cluster. As a consequence, $P_n=0$ for $n\geq 2$.
Therefore $L-N$ clusters exist, which is then also the number of
particles having exactly one hole in front. Thus we obtain
$P_{1}=(1-\rho)/\rho$ and $P_{0}=1-P_{1}=(2\rho-1)/\rho$ in this
density regime. These results are exact for $p=1$ and are reproduced
by COMF (see (\ref{P0allg})-(\ref{P0})).

The calculation of the fundamental diagram using (\ref{gfunda})
requires the knowledge of $g$. This in turn depends on the cluster
length distribution $P(k)$ which is not known exactly. Using the
approximation (\ref{gfunda}), the flow-density relation is explicitly
given by
\begin{eqnarray}
  \label{Jshuffled}
J(\rho, p=1) & = \cases{
    \rho,&  for $\rho \leq 1/2$,\\
    \frac{\rho(1-\rho)}{2\rho-1} 
    \left[\exp\left(\frac{2\rho-1}{\rho}\right)-1\right], & for $\rho>1/2$.}
\end{eqnarray}
The fundamental diagram (\ref{Jshuffled}) shows a strong asymmetry
with respect to $\rho=1/2$. For densities $\rho \leq 1/2$ each
particle can move independently and deterministically for $\rho \leq
1/2$, since every particle has at least one hole in front, exactly as
in parallel updating \cite{revisited}. If the density is increased to
values greater than $1/2$, $L-N$ clusters of nonvanishing length are
formed from which the rightmost particle can move deterministically
and since the probability to find such a particle in the system is
given by $1-\rho$, they add exactly this value to the flow, as in the
usual parallel update. Consequently, this result can be obtained from
(\ref{Jshuffled}) by a first order Taylor expansion. The contribution
of the other particles to the flow depends exponentially on the ratio
of these particles $(\rho-(1-\rho))/\rho=(2\rho-1)/\rho$.  This yields
the curvature in the fundamental diagram in the high density regime.
In contrast to the parallel and random-sequential dynamics, the
shuffled update is not particle-hole symmetric (Fig.~\ref{p1Bild}).
In this deterministic case clearly a free-flow phase ($\rho<1/2$) and
jammed phase ($\rho>1/2$) can be distinguished. They are separated by
a phase transition at $\rho_c=1/2$. The probability $P_0(\rho)$
constitutes a kind of order parameter. It vanishes exactly for
$\rho\leq 1/2$ and increases continuously for $\rho>1/2$
(Fig.~\ref{p1Bild}) indicating a second order transition. This is
similar to the deterministic limit of parallel dynamics
\cite{critical}. With decreasing hopping probability the asymmetry of
the fundamental diagram becomes smaller. In the limit $p \rightarrow
0$ we obtain the result for random-sequential update,
$J=p\rho(1-\rho)$. For $p<1$ the free-flow and jammed regimes are no
longer separated by a phase transition as can be seen from the smooth
behaviour of $P_0(\rho)$ (Fig.~\ref{p1Bild}).

We just mention \cite{woelki} that the assumption of a factorized
steady state in a finite system yields less agreement, but the
analytical results converge fast to the result of the thermodynamic
limit, derived here. In case $p=1$, for example, the largest possible
cluster has length $2N-L+1$ and the term
$\frac{\rho}{2\rho-1}\left[\exp(\frac{2\rho-1}{\rho})-1\right]$ in
(\ref{Jshuffled}) has to be replaced by the generalized hypergeometric
function $_{2}F_{2}(1, L-2N; 2, 1-N; 1)$ \cite{formel}, which
overestimates the results for small $L$, indicating the mean-field
result is not exact, at least for finite systems.


\section{Mapping onto a generalized zero-range process with parallel dynamics}
\label{GZRPsec}
The ASEP can be mapped onto a model with multi-occupation of sites,
the generalized zero-range process (GZRP). Unoccupied cells,
i.e. holes, of the ASEP become the sites of the GZRP and the particles
between the $(i-1)$-th and the $i$-th hole of the ASEP (the particles
of the $(i-1$)-th cluster) now all occupy the $i$-th site of the GZRP
and are referred to as mass $m_{i}$ (see Fig.~\ref{mapping} a) and
b)). We have $L-N$ sites and masses and still $N$ particles. At each
timestep, $l_{i}$ particles are chipped off the mass $m_{i}$ located
at site $i$ ($i=1,\ldots,L-N$) with transition probabilities
$u_{l_{i}}(m_{i})$ and are moved to site $i+1$. In the ASEP with
shuffled dynamics the probabilities $u_{l}(m)$ are given by
(\ref{transitionshuffled}). Note that although the particles in the
ASEP are updated in $N$ steps, it can be considered as if all clusters
were updated in parallel. Thus indeed the ASEP with shuffled update is
mapped onto a GZRP with parallel dynamics. In the case $p=1$ and
$N<L-N$, we have seen that in the ASEP the probability to find two
neighboring particles vanishes. For the GZRP this implies that each
cell is either occupied by exactly one particle, or it is empty. Thus
the GZRP is in fact a usual ZRP in this case and its steady state
distribution can be written as a product measure. The authors of
\cite{majumdar0} have derived a necessary and sufficient condition for
the existence of a factorized steady state distribution in a broad
class of one-dimensional mass transport models, including the
GZRP. This is possible if the transition probabilities $u_l(n)$ can be
written as a product $v_{l}w_{n-l}/\sum_{l=0}^{n}v_{l}w_{n-l}$, where
$v_{l}$ and $w_{n-l}$ are functions that depend only on $l$ and $n-l$,
respectively. They also derived a more direct test in
\cite{majumdar}. For continuous-time dynamics a related condition has
been found recently for processes in arbitrary dimensions
\cite{lebowitz}. The transition probabilities
(\ref{transitionshuffled}) of the corresponding GZRP do not satisfy
this condition, for general $p$ and $\rho$ and therefore the steady
state does not factorize, i.e.
$\tilde{P}(m_{1}, m_{2}, \dots, m_{L-N}) \not=
\tilde{P}_{m_{1}}\tilde{P}_{m_{2}}\dots \tilde{P}_{m_{L-N}}$.
In the ASEP, this implies that the particle-cluster probabilities do
not factorize. However, a factorization into hole-cluster
probabilities as assumed in COMF is not excluded.


\section{Mapping onto a zero-range process with shuffled dynamics}

\begin{figure}[!hbt]
\begin{center}
\epsfig{figure=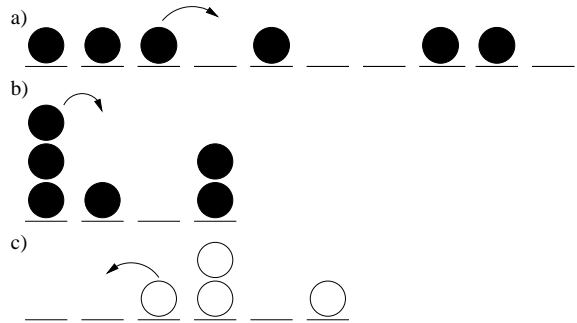,width=7.5cm,clip=}
\end{center}
\caption{\label{mapping} Mapping of the a) ASEP onto b) GZRP and c) ZRP
for $L=10$ and $N=6$. The arrows indicate a possible local transition with 
probability $u_1(3)$.}
\end{figure}

While in the GZRP an arbitrary fraction of particles is allowed to
chip off a mass, in the usual zero-range process (ZRP) \cite{spitzer}
at most one particle (say the topmost) can leave a mass during one
timestep.  The ASEP can be mapped onto a ZRP
\cite{evans_zrp,eh-review} by identifying the particles of the
exclusion model with the sites of the zero-range process, see
Fig.~\ref{mapping} a) and c). 
The holes between the $i$-th and the $(i+1)$-th particle of the ASEP
form the $i$-th mass of the ZRP. Thus the ZRP has $N$ sites and $L-N$
particles. Note that the particles in the ZRP hop to the left, as the
holes in the ASEP. While in the ASEP the particles were shuffled, in
its equivalent ZRP the sites are shuffled.  Zero-range processes are
known to have a factorized steady state distribution
\cite{evans_zrp,eh-review}. This result holds for time-continuous
dynamics and for parallel and ordered-sequential updating
\cite{evansseq} as well. For the shuffled update no general results
are available yet. The COMF approach in Sec. 3 (eq. (\ref{fact}))
describes a factorization in the ZRP-picture.  For $p=1$, the result
(\ref{P0}) is exact. However, it is not clear whether the fundamental
diagram is exact or not, since it depends also on $g$ which we do not
know exactly.

It can easily be seen that this factorization does not become exact
for general $p$, by considering the master equation for the
probability $P(0^{N-1}, M)$, i.e. one cell of the ZRP having
occupation number $M:=L-N$, followed by $N-1$ empty sites (a cluster of N
particles followed by $L-N$ holes in the ASEP):
\begin{equation}
\fl P(0^{N-1}, M) = \left(\bar{p} + \frac{p^{N}}{N!} \right) P(0^{N-1}, M)
+ \bar{p} \sum\limits_{k=1}^{N-1}\frac{p^{k}}{k!} P(0^{N-k-1}, 1,
0^{k-1}, M-1),
\end{equation}
with $\bar{p}=1-p$. Inserting the product measure ansatz (\ref{fact})
and performing the limit $N \rightarrow \infty$ gives 
\begin{equation}
\frac{p}{\bar{p}} =
\frac{P_{1}}{P_{0}}\frac{P_{M-1}}{P_{M}} \left(e^{p}-1 \right).
\end{equation}
Now, inserting the expression for $P_{n}$, $n\geq 1$ from
eq. (\ref{Pnallg}) and using (\ref{gshuffled}) leads to the constraint
\begin{equation}
\label{cond}
0 = (e^{p}-1)\left(\frac{p}{e^{pP_{0}}-1} - \frac{1-P_{0}}{P_{0}} \right) - p.
\end{equation}
This condition is only fulfilled in the limit $P_{0}=1$ (i.e. for
$\rho=1$) and also for $p=0$. Denoting the r.h.s. of (\ref{cond}) by
$\xi(p, P_{0})$ one can see that $\xi$ increases almost linearly for
decreasing $P_{0}$. In the limit $P_{0} \rightarrow 0$ it becomes
maximal, taking the value $\xi(p, 0) = ((2-p) e^p - 2 - p)/2p$, which
is, for $p \not= 0$ distinctly different from zero. Thus one can
conclude that the COMF theory gives a good approximation but not the
exact result, for general $p$. In other words, the ZRP with shuffled
update does not have a product measure steady state.


\section{Truncated processes}
\label{trunc-sec}
In Sec.~\ref{COMFsection} we have seen through mapping onto a
GZRP that the shuffled update can also be interpreted as a
cluster dynamics. The probabilities that $l$ particles leave a
cluster of length $n$ are given by (\ref{transitionshuffled}).
These probabilities decrease rapidly with increasing $l$. Therefore
it is natural to consider as an approximation to the
original dynamics models in which, irrespective of the cluster length,
at most $l_{\rm max}$ particles are allowed to leave a cluster 
at any given timestep. This truncation leads to a GZRP defined
by probabilities
\begin{equation}
\label{transitiontruncated}
u^{(l_{\rm max})}_{l}(n) = \frac{p^{l}}{l!} - 
\frac{p^{l+1}}{(l+1)!}\theta(l_{\rm max}-l),
\end{equation}
with $0 \leq l \leq \min(l_{\rm max},n)$ (and $0$ otherwise). The
case $l_{\rm max}=1$ corresponds to the standard ZRP with hopping
probability $u(n)=p$, i.e.\ the ASEP with parallel dynamics.  Thus, the
fundamental diagram is given by $J_1(\rho,
p)=\frac{1}{2}\left(1-\sqrt{1-4 p \rho(1-\rho)}\right)$ \cite{ito}.

\begin{figure}[!hbt]
\begin{center}
\epsfig{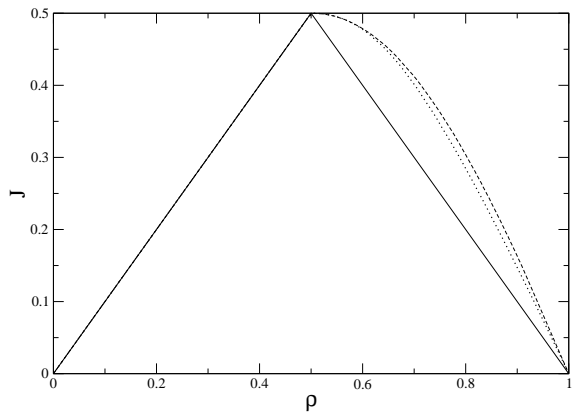}
\end{center}
\caption{\label{zwei} Comparison of truncated models for $p=1$. The
continuous curve shows the case $l_{{\rm max}}=1$, i.e. the case of
usual parallel update. The dotted curve shows $l_{\rm max}=2$, given
by (\ref{J2}) and the dashed curve shows the fundamental diagram for
the shuffled update, obtained for $l_{\rm max}=\infty$.}
\end{figure}

In the case $l_{\rm max}=2$ the factorization condition of
\cite{majumdar0,majumdar} is satisfied for $p=1$ and we have
$\tilde{P}(m_{1}, m_{2}, \dots,
m_{L-N})=\tilde{P}_{m_{1}}\tilde{P}_{m_{2}}\dots \tilde{P}_{m_{L-N}}.$
For the fundamental diagram we obtain
\begin{eqnarray}
  \label{J2}
J_{2}(\rho, p=1) & = \cases{
    \rho,&  for $\rho \leq 1/2$,\\
    1- \frac{\rho + \sqrt{\rho^2 - 2(1-\rho)(2\rho-1)}}{2}, & for $\rho>1/2$.}
\end{eqnarray}
Fig.~\ref{zwei} shows the results for $l_{\rm max}=1,2$ in comparison
with the shuffled update $l_{\rm max}=\infty$. The steady state
probabilities for $l_{max}=2$ depend on the number of
particle-clusters with length $1$. This is not correctly reproduced by
COMF theory which assumes a factorization into hole-cluster
probabilities (\ref{fact}) and yields that all stationary
configurations for fixed particle number $N$ are equally probable for
$p=1$ and $\rho>1/2$: $P(n_{1}, n_{2}, \dots,
n_{N})=P_{0}^{2N-L}P_{1}^{L-N}$. Therefore one can conclude that the
steady state distribution for the truncated model with $l_{max}=2$ and
$p=1$ factorizes into particle-cluster probabilities but not into
hole-cluster probabilities.

For $l_{\rm max}=3,4,\dots$ the curves converge quickly to the
shuffled curve, but they no longer fulfill the factorization
criterion. 

\section{Application to pedestrian dynamics}
\begin{center}
\begin{figure}[!ht]
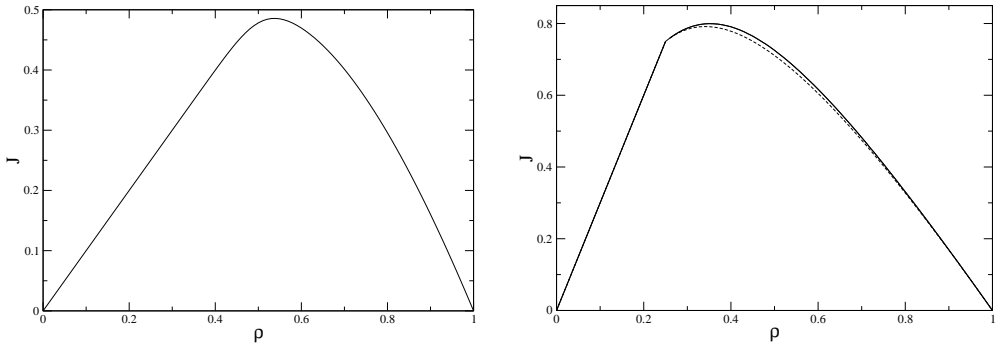

\centerline{\epsfig{figure=shuff_encoupled_L100_L10.eps, height=4.5cm,clip=} 
\quad\epsfig{figure=vgl_vmax3.eps, height=4.5cm, clip=}}
  \caption{Fundamental diagrams for two generalized processes with
    $p=1$.  The left diagram shows the result for the model with
    decoupled lanes for $W=10$. The right one shows the case of
    increased velocity $v_{{\rm max}}=3$ for $W=1$. The dashed curve
    represents the analytical result, the continuous curve the result
    from computer simulations.}
  \label{v}
\end{figure}
\end{center}

As mentioned in the beginning, the shuffled update has an
interpretation \cite{kluepfel,kluepfeldiss} as an update procedure in
two-dimensional cellular automata to avoid conflicts (situations in
which more than one particle tries to access the same cell) occurring
in parallel dynamics \cite{friction}. An important example
is the movement of pedestrians in a long corridor.  To describe this 
movement more realistically we consider two straightforward generalizations 
of the model in the following. The simplest extension of the
ASEP, which corresponds to a single-lane model is a model of several 
decoupled lanes. Here pedestrians are not allowed to change lane, i.e.\
their $y$-coordinates are fixed.
If in the initial state the pedestrians are distributed 
stochastically on sites (and thus lanes), their density in
each lane is fixed and constant over time. 
The probability of finding $i$ pedestrians in a
particular lane is given by the hypergeometric distribution 
\begin{equation}
{\rm Hyp}_{N,L,L(W-1)}(i) = {L \choose i}{L(W-1)
\choose N-i}/{LW \choose N}
\end{equation} 
with mean $\rho=N/(LW)$. The fundamental diagram of the model
with decoupled lanes is \cite{burstedde}
\begin{equation}
\tilde{J}(\rho) = \sum_{i=0}^{L} {\rm Hyp}_{\rho LW, L, L(W-1)}(i)
\cdot J(i/L), 
\end{equation}
where $J(\rho)$ is the single-lane fundamental diagram 
determined in Sec.~\ref{COMFsection}. Using the approximation
(\ref{gfunda}), the result for a width of $W=10$
cells and deterministic hopping ($p=1$) in the thermodynamic limit is
depicted in Figure \ref{v} (a). One can see that the maximum is
smoother and a little bit lowered in comparison to the corresponding
Figure \ref{p1Bild} (b). Note that lane-changing, an effect that is
especially relevant for high densities, would lead to a reduction of
the flow.
To reproduce the shift of the maximum flow to lower densities,
observed in pedestrian dynamics, we introduce a larger maximum
velocity $v_{{\rm max}}$, i.e. a larger number of cells that can be
passed during one timestep \cite{kknss}.  The simplest way is to
allow a pedestrian to move the minimum of $v_{{\rm max}}$ cells and
the number of empty cells in front with probability $p$ 
at each timestep. In the following we again consider the case $W=1$
and $p=1$ only.  For densities less or equal to $1/(v_{{\rm max}}+1)$
each of the pedestrians has at least $v_{{\rm max}}$ empty sites in
front, i.e. the probability to find a particle in front is $P_{0}=0$
and the flow is determinstically given as $v_{{\rm max}}\rho$.  For
higher densities it may happen that e.g. a pedestrian occupies a cell
at time $t+1$ that has been occupied by a different pedestrian
belonging to the cluster in front at time $t$. However we found in
computer simulations that this effect can be neglected (for relatively
small $v_{\rm max}$). Thus one can again describe the dynamics in the
picture in which all clusters are updated in parallel. The system
tries to separate the clusters by exactly $v_{\rm max}$ holes. The
probability to have at least one hole in front is in this case
$(1-\rho)/v_{\rm max}$ and the ratio of particles which do not have a
particle in front is $(\rho - (1-\rho)/v_{\rm max})/\rho$. This
defines the probability $P_{0}$, which can finally be written as
$\left[(v_{\rm max}+1)\rho-1\right]\theta(\rho-1/(v_{\rm
max}+1))/(v_{\rm max}\rho)$. With the use of (\ref{Jshuffled}) we
obtain
\begin{eqnarray}
  \label{Jshuffledvmax}
\fl J(\rho, v_{\rm max}, p=1) & = \cases{ v_{\rm max}\rho, &for $\rho
    \leq 1/(v_{\rm max}+1)$,\\ \frac{v_{\rm
    max}\rho(1-\rho)}{(1+v_{\rm max})\rho-1}
    \left[e^{\frac{(v_{\rm max}+1)\rho-1}{v_{\rm
    max}\rho}}-1\right], & else.}
\end{eqnarray}
In Figure \ref{v} (b) the resulting fundamental diagram in case of
$v_{{\rm max}}=3$ and $p=1$ is depicted. The dashed line shows the
analytical in comparison to the numerical result.  In opposite to the
case of maximum velocity $1$, depicted in Figure \ref{p1Bild} (b), the
maximum of the fundamental diagram and the critical point $(1/(v_{\rm
max}+1), \rho v_{\rm max})$ do not coincide; this is qualitatively
reproduced by the analytical result. However, to recover the exact
result one would have to take longer ranged correlations into account.

\section{Discussion}
In this paper the asymmetric exclusion process (ASEP) with periodic
boundary conditions and shuffled dynamics was studied. Using a
mean-field approach, steady state properties like the fundamental
diagram and distribution functions are derived approximatively that
are in very good agreement with data from Monte-Carlo simulations of
large systems.


The model can be mapped onto a generalized zero-range process (GZRP)
with usual parallel update, by interpreting the holes of the ASEP as
the sites of the GZRP. These sites are occupied by the particles which
follow the corresponding holes of the ASEP to the left. Further, the
model can be mapped onto a zero-range process (ZRP) with shuffled
update by interpreting the particles of the ASEP as sites of the
ZRP. These sites are occupied by particles, that correspond to the
holes of the ASEP.

In the case $p=1$ the exact results for the steady state distribution
were presented. For densities $\rho < 1/2$ all particles are separated
by at least one hole and thus the corresponding GZRP factorizes (see
sec. \ref{GZRPsec}). Every configuration $\tilde{P}(m_{1}, \dots,
m_{L-N})$ can thus be written as a product
$\tilde{P}_{1}^{N}\tilde{P}_{0}^{L-2N}$.  For $\rho>1/2$, no
neighboring holes occur in the ASEP and thus in the corresponding ZRP
sites are either occupied by exactly one particle or are empty with
probabilities $P_{1}=(1-\rho)/\rho$ and $P_{0}=(2\rho-1)/\rho$,
respectively.
However, since the shuffled dynamics is intrinsically stochastic,
already for $p=1$ the fundamental diagram is nontrivial and
asymmetric. For $\rho \leq 1/2$ all particles can move
deterministically and the stationary flow is identical to the
deterministic limit of the parallel update which is known exactly. For
higher densities $\rho>1/2$ the probability for a particle to move
depends on the number of particles in front and therefore the flow
depends on the cluster-size distribution. However, mean-field theory
yields a good approximation for the flow.
The two regimes are separated by a second order phase transition. In
the case $p<1$ the two regimes can no longer be distinguished and the
flow is determined by the stochasticity for all densities. The
fundamental diagrams become smooth, but still do not exhibit a
particle-hole symmetry. The free flow and jammed regimes are no longer
separated by a phase transition.  It could be shown that in this more
general case neither the GZRP, nor the ZRP have a factorized steady
state. However, the mean-field theory yields again a very good
approximation. In the limit $p \rightarrow 0$ the model can be
described by a product measure since the steady state distribution
converges to the result of random-sequential updating.


In contrast to the other updates investigated so far it is essential
that we are dealing with distinguishable particles. The numbering is
needed in order to define the update. This is not the case for
random-sequential, parallel or ordered-sequential updates. Note that
in the random-sequential case sites could be updated instead of
particles without changing the stationary state. For the
ordered-sequential dynamics only one particle has to be marked, namely
the first one of the sequence.

A truncated process in which at most two particles of a cluster can
move and its exact solution for $p=1$ was presented. It could be shown
that this model factorizes into particle-cluster probabilities but not
into hole-cluster probabilities. Since in shuffled dynamics the
probability that the third particle in a cluster moves is relatively
small, the truncated dynamics gives also a good approximation for the
model with shuffled update.

It has been shown \cite{kluepfel,kluepfeldiss} that cellular automata
with shuffled update have an interpretation as pedestrian flow
models. In this sense the ASEP with shuffled update models the
directed motion of pedestrians in a long corridor of small width. To
allow pedestrians to walk side by side, we generalized the model to a
two-dimensional scenario with decoupled lanes, in which pedestrians
can not change lanes. Further a generalization to higher maximum
velocities was given. For both generalizations approximations were
presented that are in good agreement with Monte-Carlo data.

We want to mention that the idea of ordered sequential updates
(forward or backward ordered sequences) can be generalized
\cite{woelkiprep}: Consider a general ordered sequential update with
sequence ${\bf \pi} = (\pi(1), \pi(2),\dots, \pi(N))$, which now is
quenched for all timesteps. $\pi(1)$ denotes the update number of the
particle with number $1$ and so on, where the particles are numbered
from left to right.  The forward sequential dynamics is the special
case of $\pi_{F} = (1, 2,\dots, N )$ and the backward sequential
dynamics is the special case of $\pi_{B} = (N, N-1,\dots, 1 )$. As far
as we know, other update sequences have not been studied yet. However,
it is easy to show that for $N$ even all $\left((N/2)!\right)^2$
sequences, in which the even (odd) particles are updated first, and
then the odd (even) particles, have a factorized steady state
distribution \cite{woelki}. Note that these updates are equivalent to
a sublattice-parallel-like update, in which in the first half of a
timestep all even (odd), and in the second half, all odd (even)
particles are updated in parallel.  However, the usual ordered
sequential update does not factorize with a general update
sequence. Hence, it would be interesting to find a condition for a
model with sequence $\pi$ (and transfer matrix $h^{(1)}\dots h^{(N)}$)
to have a factorized steady state. Among that, if an update with
sequence $\pi$ generates the steady state $F(\pi)$, it is interesting
to consider the average of the steady states generated by all
permutations of the permutation group $S$: $\langle F \rangle =
\sum_{\pi \in S} F(\pi)/N!$.  We found in computer simulations that
this steady state differs from the steady state of the shuffled update
distinctly, what can not be trivially anticipated.

Summarizing, maybe the most surprising result is the fact that the
ASEP with shuffled update does not factorize 
in contrast to the updates investigated before. Here
the distinguishibility of the particles seems to be important whereas
in the updates considered previously the particles are basically
indistinguishable. This points warrants further investigations.\\

\vspace{0.7cm}
\noindent
{\bf Acknowledgements}\\
\noindent
We thank Ansgar Kirchner and Hubert Kl\"upfel for helpful
discussions.

\end{document}